%% file: main.tex
\documentclass[acmtog,nonacm]{acmart}
\usepackage[ruled,vlined,linesnumbered]{algorithm2e}
\usepackage{amsfonts}
\usepackage[utf8]{inputenc}
\usepackage{makecell}
\usepackage{colortbl}
\definecolor{highlightgreen}{HTML}{46D6EB}

\SetAlFnt{\small}
\SetAlCapFnt{\small}
\SetAlCapNameFnt{\small}
\SetAlCapHSkip{0pt}

\author{Laurent Vit}
\affiliation{%
  \institution{University of Canterbury}
  \country{New Zealand}
}

\author{Oliver Batchelor}
\affiliation{%
  \institution{University of Canterbury}
  \country{New Zealand}
}

\author{Richard Green}
\affiliation{%
  \institution{University of Canterbury}
  \country{New Zealand}
}

\keywords{Gaussian Splatting, Ray Tracing, Optimization, Radiance Fields}
\title{3D Gaussian Accelerated Ray Tracing: Fast training through particle-based backward propagation}

\acmSubmissionID{2266}
\begin{document}
\begin{teaserfigure}
\includegraphics[width=\textwidth]{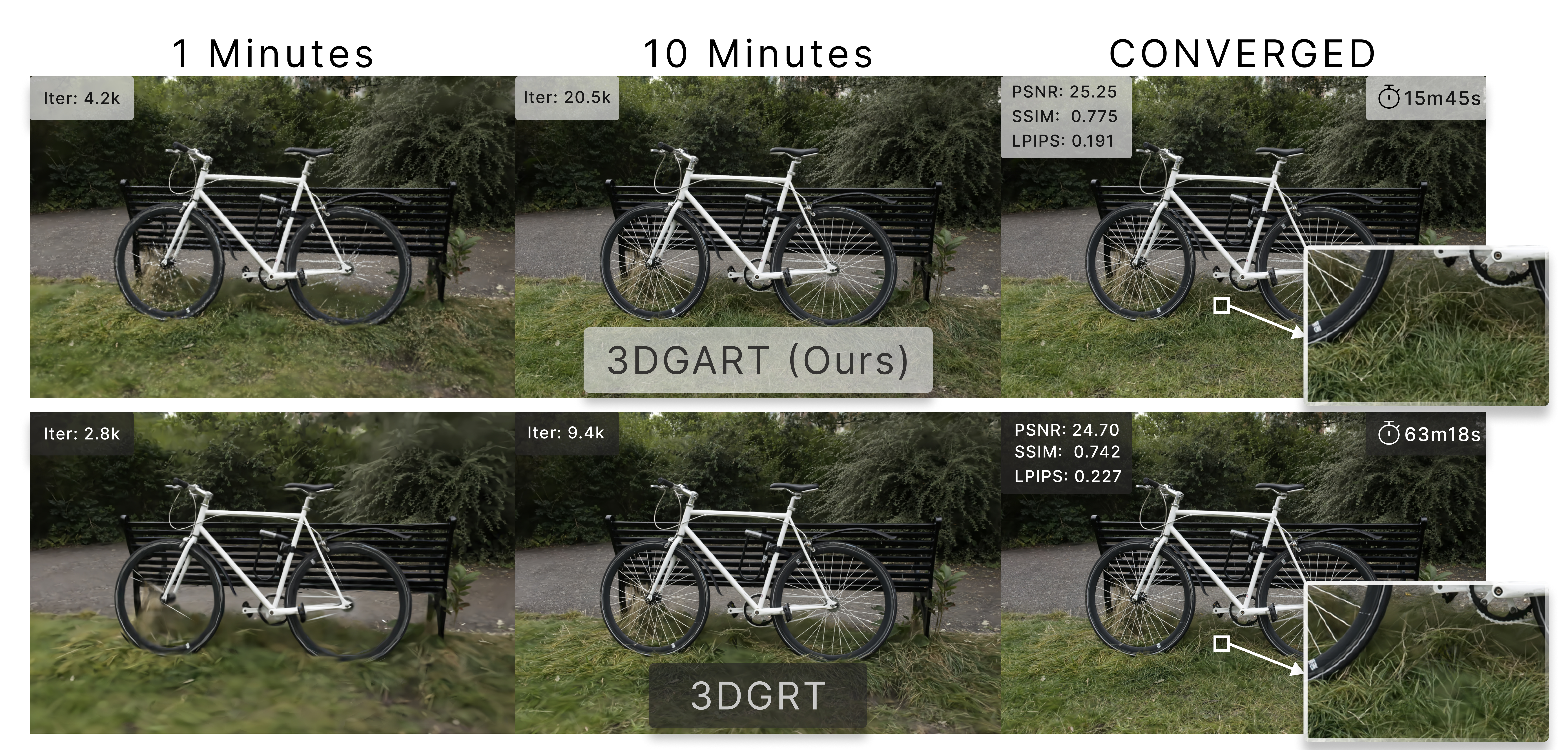}
\caption{
NVIDIA 3DGRT improves efficiency through compromises, including
a shorter Gaussian kernel, a stricter opacity cutoff, and capped
ray-primitive intersections. In contrast, 3DGART (Ours) preserves the ray-traced
Gaussian formulation and accelerates training by reorganizing the
backward pass. On the Mip-NeRF 360 \emph{bicycle} scene, 3DGART converges \textbf{$\approx4\times$} faster than 3DGRT, while achieving higher
reconstruction quality.
}
\label{fig:teaser}
\end{teaserfigure}

\input{Chapters/abstract.tex}
\maketitle

\input{Chapters/1_introduction.tex}
\input{Chapters/2_related_work.tex}
\input{Chapters/3_background.tex}
\input{Chapters/4_methods.tex}
\input{Chapters/5_experimentation.tex}
\input{Chapters/6_discussion.tex}
\input{Chapters/7_conclusion.tex}
\input{Chapters/acknowledgement}

\bibliographystyle{ACM-Reference-Format}
\bibliography{references}

\end{document}

%% file: Chapters/abstract.tex
\begin{abstract}3D Gaussian Splatting has made Gaussian primitives a highly efficient representation for real-time novel view synthesis, but its rasterisation-based formulation relies on screen-space approximations that limit accurate view-dependent ordering and the integration of secondary ray effects such as reflections, refractions, and shadows. Gaussian ray tracing addresses these limitations by evaluating explicit ray-primitive intersections, yet it remains costly to train. We observe that the main bottleneck is not ray traversal alone, but the pixel-centric backward propagation, where many threads concurrently accumulate gradients into the same primitive parameters, causing severe atomic contention and thread serialisation.

We present 3DGART, a practical training framework for ray-traced Gaussian rendering. Our key idea is to reorganise backward propagation around primitives rather than pixels. Using conservative perspective-correct screen-space bounds, we build a compact intermediate buffer and a tile-primitive mapping that allows each thread to accumulate the contribution of one primitive over its covered pixels within a tile. This transforms gradient computation from a contention-heavy scatter operation into a structured gather-like process. On Mip-NeRF 360, 3DGART achieves an $\approx 3-3.5\times$ raw training speedup over per-pixel baseline and $\approx4 \times$ over 3DGRT on Mip-NeRF 360 while improving quality. More importantly, 3DGART makes fully ray-traced Gaussian training practical, reaching runtimes competitive with rasterisation-based pipelines while preserving benefits of ray tracing.
\end{abstract}

%% file: Chapters/1_introduction.tex
\section{Introduction}

3D Gaussian Splatting (3DGS) \cite{kerbl_3d_2023} has rapidly become one of the dominant representations for real-time novel view synthesis. By representing a scene as a set of anisotropic Gaussian primitives and projecting them to screen space, 3DGS combines high visual fidelity with extremely efficient rendering. This efficiency has made Gaussian splatting attractive for reconstruction, editing, and real-time applications. However, its performance is largely enabled by a rasterisation pipeline: Gaussians are approximated in screen space, sorted in image-space tiles, and composited using projective footprints. While highly effective, this formulation inherits fundamental limitations from screen-space rendering, including projection inaccuracies \cite{zwicker_ewa_2001}, view-dependent sorting artefacts, aliasing, and limited support for secondary rays or non-standard camera models.

Gaussian Ray tracing \cite{moenne-loccoz_3d_2024} offers a conceptually cleaner alternative. Instead of approximating Gaussian influence through screen-space splats, ray-traced Gaussian methods evaluate explicit ray-primitive interactions. This formulation naturally supports perspective-correct visibility, complex camera models, and more general light transport effects. Recent systems such as 3D Gaussian Ray Tracing demonstrate that Gaussian radiance fields can be rendered through hardware-accelerated ray tracing, opening the door to a unified representation for primary visibility, reflections, refractions, and other ray-based effects. Yet despite these advantages, ray-traced Gaussian methods have not displaced rasterisation in practice. The reason is simple: training remains too slow.

In this paper, we argue that the main obstacle to practical ray-traced Gaussian training is not only ray traversal, but the structure of the backward pass. Existing differentiable ray-tracing pipelines follow a pixel-centric formulation: each thread processes a ray, evaluates the primitives intersected by that ray, and accumulates gradients into the corresponding Gaussian parameters. Since the same primitive may be visible in many pixels, many threads concurrently update the same memory locations. This results in heavy use of atomic operations, severe contention, and thread serialisation. As the number of primitives and ray-primitive interactions grows, the backward pass becomes the major computational bottleneck. Existing systems mitigate this issue through practical restrictions such as modified Gaussian kernels, $\alpha\_{min}$ or capped ray intersections, but these heuristics trade accuracy and scalability for speed rather than addressing the root cause.

A particle-centric formulation offers a more suitable organisation for training. Instead of assigning ownership to pixels and scattering gradients into shared primitive parameters, gradients can be accumulated by assigning ownership to primitives over local regions of the image. Such an organisation largely avoids fine-grained atomic contention and better matches the structure of the scene representation. However, making this idea practical is non-trivial: the backward pass must efficiently determine which pixels are affected by each primitive, recover the intermediate values required for differentiation, access them with coherent memory patterns on the GPU and consume the least possible memory.

We introduce 3D Gaussian Accelerated Ray Tracing (3DGART), a framework designed to make ray-traced Gaussian training practical. Our central idea is to replace the conventional pixel-centric backward pass with a particle-centric formulation. Instead of assigning gradient accumulation to pixels that scatter updates into shared primitive parameters, we organise computation around primitives within local image-space tiles. This changes the structure of differentiation: each thread is responsible for accumulating the contribution of one primitive over the pixels it covers inside a tile. The backward pass is therefore transformed from a contention-heavy scatter operation into a structured gather-like computation with greatly reduced atomic pressure.

Together, these choices lead to the following contributions:

\begin{itemize}
    \item We identify atomic contention in the pixel-centric backward pass as the primary bottleneck preventing practical ray-traced Gaussian training.
    \item We propose a compact intermediate-buffer layout based on perspective-correct projected bounds, enabling efficient storage and retrieval of the values required for backpropagation.
    \item We introduce a particle-centric backward propagation scheme based on a tile-primitive mapping, substantially reducing gradient accumulation contention.
    \item We demonstrate that 3DGART achieves up to $4\times$ training speed-up over 3DGRT and brings ray-traced Gaussian training into the runtime range of rasterisation-based methods.
\end{itemize}

%% file: Chapters/2_related_work.tex
\section{Related Work}

\subsection{Neural Radiance Fields}
Neural Radiance Fields (NeRF) \cite{mildenhall_nerf_2020} have brought about a fundamental shift in novel view synthesis, representing scenes as continuous volumetric functions encoded within a multi-layer perceptron (MLP). Subsequent work has driven significant progress in rendering quality \cite{barron_mip-nerf_2022, barron_zip-nerf_2023} and training speed \cite{muller_instant_2022}. However, the computational cost inherent to neural networks remains prohibitive for real-time operation, and the implicit nature of NeRF representations complicates direct scene manipulation. These limitations have motivated the shift toward explicit, particle-based representations.

\subsection{Point-Based Radiance Fields \& Gaussian Splatting}

3D Gaussian Splatting (3DGS) \cite{kerbl_3d_2023} has quickly established itself as the dominant paradigm for real-time novel view synthesis. Rather than sampling along rays, 3DGS projects anisotropic Gaussian primitives directly onto the image plane via EWA splatting \cite{zwicker_ewa_2001}, sorting them within discrete screen-space tiles for efficient alpha-compositing. This tile-based rasterisation pipeline achieves exceptional throughput and has catalysed a broad research effort, such as quality improvements \cite{kheradmand_3d_2025, liu_deformable_2025}, execution speed \cite{feng_flashgs_2024, hahlbohm_faster-gs_2026}, and model compression \cite{mallick_taming_2024, hanson_speedy-splat_2025}.

Despite its efficiency, rasterised Gaussian splatting remains tied to a screen-space approximation of the underlying 3D primitives. Projected footprints, depth ordering based on primitive centres, and tile-local sorting can introduce view-dependent artefacts, popping \cite{radl_stopthepop_2024}, and aliasing \cite{yu_mip-splatting_2023}. Recent hybrid approaches improve perspective correctness \cite{hahlbohm_efficient_2025} and support distorted cameras \cite{wu_3dgut_2025, huang_3dgeer_2026} within rasterisation pipelines. However, they still inherit the core design of screen-space splatting, motivating the need to directly evaluate Gaussian contribution along rays.

\subsection{Differentiable Ray Tracing}

Recent works have explored this direction through fully ray-traced Gaussian representations. 3D Gaussian Ray Tracing (3DGRT)~\cite{moenne-loccoz_3d_2024} replaces projective splats with explicit ray--primitive intersections, representing Gaussian support with bounded geometry inside a hardware ray-tracing acceleration structure. EVER \cite{mai_ever_2025, blanc_raygauss_2025, blanc_raygaussx_2025} further improves the quality by performing exact volumetric integration or sampling densities along rays.

Concurrent and recent systems \cite{lee_grtx_2026} orbit around improving the ray-tracing back-end, including acceleration structure design, primitive representation \cite{condor_dont_2024}, and traversal efficiency. While these contributions reduce the computational cost of the forward pass, Gray \cite{poirier-ginter_gray_2026} instead alleviates performance bottlenecks through a dense initialization of narrow Gaussians, combined with few compromises, representing an accuracy--performance trade-off. Ultimately, existing methods either optimize traversal efficiency or bypass the main bottleneck: the backward pass.
 
In differentiable ray-traced Gaussian pipelines, the backward pass remains pixel-centric: each ray thread scatters gradient contributions into shared primitive parameters, producing heavy atomic contention when many pixels observe the same primitive.

%% file: Chapters/3_background.tex
\section{Background}

\subsection{Gaussian Scene Representation}

We represent a scene as a set of anisotropic 3D Gaussian primitives. Each primitive $i$ is defined by a mean position $\boldsymbol{\mu} \in \mathbb{R}^3$ and a covariance matrix $\boldsymbol{\Sigma} \in \mathbb{R}^{3 \times 3}$, which encodes its spatial extent and orientation. The density response of a Gaussian at a 3D point $\mathbf{x} \in \mathbb{R}^3$ is given by
\begin{equation}
    \rho(\mathbf{x}) =
    \exp \left(
        -\frac{1}{2}
        (\mathbf{x} - \boldsymbol{\mu})^\top
        \boldsymbol{\Sigma}^{-1}
        (\mathbf{x} - \boldsymbol{\mu})
    \right).
\end{equation}

Following 3D Gaussian Splatting \cite{kerbl_3d_2023}, the covariance is parameterized as
\begin{equation}
    \boldsymbol{\Sigma} =
    \mathbf{R} \mathbf{S} \mathbf{S}^\top \mathbf{R}^\top,
\end{equation}
where $\mathbf{R} \in \mathbb{R}^{3\times3}$ is a rotation matrix and $\mathbf{S} \in \mathbb{R}^{3\times3}$ is a diagonal scaling matrix. This parameterisation ensures that the covariance remains positive semi-definite during optimisation.

Each primitive also stores an opacity $\sigma_i$ and a view-dependent radiance function $\mathbf{c}_i(\mathbf{d})$, typically represented using spherical harmonics, where $\mathbf{d}$ denotes the viewing direction.

\subsection{Ray-Gaussian Evaluation}

Given a camera ray
\begin{equation}
    \mathbf{r}(\tau) = \mathbf{o} + \tau \mathbf{d},
\end{equation}
Ray-traced Gaussian rendering evaluates the contribution of each intersected primitive along the ray. Following 3DGRT \cite{moenne-loccoz_3d_2024}, this evaluation is performed in the normalised local space of the Gaussian, where the transformed primitive corresponds to a unit sphere. Therefore, the point of maximum Gaussian response along the ray is:
\begin{equation}
    \tau_{\max}
    =
    -
    \frac{
        \mathbf{o}_g^\top \mathbf{d}_g
    }{
        \mathbf{d}_g^\top \mathbf{d}_g
    }.
\end{equation}
where
\begin{equation}
    \mathbf{o}_g = \mathbf{S}^{-1}\mathbf{R}^\top(\mathbf{o} - \boldsymbol{\mu}),
    \qquad
    \mathbf{d}_g = \mathbf{S}^{-1}\mathbf{R}^\top\mathbf{d}.
\end{equation}

\subsection{Alpha Compositing}

For a sorted sequence of ray-primitive interactions, rendering is performed using front-to-back alpha compositing. The final pixel colour is
\begin{equation}
    \mathbf{C}
    =
    \sum_{i=1}^{N}
    T_i \alpha_i \mathbf{c}_i(\mathbf{d}),
\end{equation}
where
\begin{equation}
    T_i =
    \prod_{j<i} (1 - \alpha_j)
\end{equation}
is the accumulated transmittance before the $i$-th primitive.

The opacity contribution of primitive $i$ is defined as
\begin{equation}
    \alpha_i =
    \sigma_i \rho_i(\mathbf{x}_i),
\end{equation}
where $\mathbf{x}_i$ denotes the maximum-response point of the Gaussian along the ray.

\subsection{Ray Tracing scene representation}

Modern hardware-accelerated ray tracing systems represent scenes as collections of geometric primitives organised within hierarchical acceleration structures, enabling efficient ray-primitive intersection queries. In particular, RT cores are optimised for triangle intersections \cite{condor_dont_2024}.

Recent Gaussian ray-tracing systems, such as GRTX \cite{lee_grtx_2026}, follow a two-level acceleration hierarchy. A bounded primitive (e.g., an icosphere) is stored once in a bottom-level acceleration structure (BLAS). At the same time, each Gaussian is represented as an instance in a top-level acceleration structure (TLAS), varying its position, orientation, and scale through per-instance transforms.

The mapping from the unit primitive to world space is defined by the affine transform
\begin{equation}
\mathbf{M} =
\begin{bmatrix}
m_{00} & m_{01} & m_{02} & \mu_x \\
m_{10} & m_{11} & m_{12} & \mu_y \\
m_{20} & m_{21} & m_{22} & \mu_z
\end{bmatrix}
=
\left[
\mathbf{R} \mathbf{S} \alpha_{c}
\mid
\boldsymbol{\mu}
\right],
\end{equation}
where $\mathbf{R} \in \mathbb{R}^{3 \times 3}$ respectively represent the rotation and scaling matrix, $\mathbf{S} \in \mathbb{R}^{3 \times 3}$, $\boldsymbol{\mu}$ is its centre, and
\begin{equation}
    \alpha_{c}
    =
    \sqrt{
        2 \log
        \left(
            \frac{\sigma}{\alpha_{\min}}
        \right)
    }
\end{equation}
is an opacity-dependent clamping radius which follows \cite{radl_stopthepop_2024}. Typically, we use $\alpha_{\min}=1/255$.

\subsection{Pixel-Centric Backpropagation Bottleneck}

Differentiable ray-traced Gaussian renderers typically parallelise computation over pixels. While this mapping is natural for the forward pass, it becomes inefficient during backpropagation. Since multiple pixels observe the same primitive, many threads concurrently update shared parameters, leading to severe atomic contention and thread serialisation.

As a result, any gain in training time, even through efficient ray traversal, is absorbed by gradient accumulation. This limitation motivates a different organisation of the backward pass, where gradients are accumulated per primitive rather than per pixel. The challenge is to construct this mapping efficiently while preserving coherent GPU execution, which we address in the following section.

%% file: Chapters/4_methods.tex
\section{METHOD}

\subsection{Overview}

Our goal is to make ray-traced Gaussian training practical by reorganising the backward pass from a pixel-centric scatter operation into a primitive-centric, gather-like computation. The forward pass remains ray-traced: visibility, ordering, and ray-Gaussian interactions are still determined by hardware ray tracing. The main challenge is therefore not to approximate these interactions, but to efficiently determine which pixels must be revisited for each primitive and to retrieve the intermediate quantities required for differentiation. 3DGART addresses this challenge in three steps:
\begin{itemize}
    \item First, we compute conservative perspective-correct screen-space bounds for each Gaussian, which define the set of pixels that may interact with the primitive.
    \item Second, we use these bounds to allocate a compact intermediate buffer storing the accumulated colour and transmittance values required for backpropagation.
    \item Third, we build a tile-primitive mapping that allows the backward pass to assign work to primitive-tile pairs.
\end{itemize}

\subsection{Perspective-correct Footprint \& Memory Pre-Allocation}
\paragraph{Perspective-correct Gaussian bounds.}
We first compute a conservative perspective-correct screen-space AABB for each ray-traced Gaussian proposed by \cite{weyrich_hardware_2007}. This bound identifies the pixels whose rays may intersect the primitive, while remaining consistent with the affine transform used by the ray tracer.
Let
\begin{equation}
    \mathbf{T}' = \mathbf{V}\mathbf{P}\mathbf{M}\mathbf{T},
    \qquad
    \mathbf{T}
    =
    \begin{bmatrix}
        \mathbf{R}\mathbf{S} & \boldsymbol{\mu} \\
        \mathbf{0} & 1
    \end{bmatrix}.
    \label{eq:projected_transform}
\end{equation}

where $\mathbf{VPM} \in \mathbb{R}^{4\times4}$ is the camera
transformation matrix that maps world-space coordinates into clip
space, allowing the Gaussian primitive to be projected onto the image
plane. \(\mathbf{T}\in\mathbb{R}^{4\times4}\) maps the normalized Gaussian space to the world space, \(\mathbf{R}\mathbf{S}\in\mathbb{R}^{3\times3}\) contains the oriented support axes of the Gaussian, and \(\boldsymbol{\mu}\in\mathbb{R}^3\) is its center. ~\cite{hahlbohm_efficient_2025} scales the projected ellipsoid by an opacity-dependent cut-off \(\rho_c\), which is already absorbed into our affine transform \(\mathbf{R}\mathbf{S}\). The interval along each axis is then:
\begin{equation}
    [b_i,t_i] = [p_i-h_i,\,p_i+h_i].
    \label{eq:screen_bounds}
\end{equation}
where
\begin{equation}
    \mathbf{v} = (1,1,1,-1),
    \qquad
    s =
    \left\langle
        \mathbf{v},
        \mathbf{T}'_4 \odot \mathbf{T}'_4
    \right\rangle,
    \qquad
    \mathbf{f} = \frac{\mathbf{v}}{s}.
    \label{eq:quadric_factor}
\end{equation}

\begin{equation}
    p_i =
    \left\langle
        \mathbf{f},
        \mathbf{T}'_i \odot \mathbf{T}'_4
    \right\rangle,
    \qquad
    h_i =
    \sqrt{
        p_i^2 -
        \left\langle
            \mathbf{f},
            \mathbf{T}'_i \odot \mathbf{T}'_i
        \right\rangle
    }.
    \label{eq:center_radius}
\end{equation}

\paragraph{Memory Preallocation}The aggregate footprint of all projected primitives then gives an upper bound on the required intermediate storage. The capacity is then:

\begin{equation}
|\mathcal{LT}| = \lambda_s \, C \sum_{i=1}^{N} A_i
\label{eq:lt_buffer_capacity}
\end{equation}

\noindent
where $N$ is the number of Gaussian primitives, $C=4$ is the number of stored
floating-point channels per entry, corresponding to the RGB accumulated colour
$L \in \mathbb{R}^3$ and transmittance $T \in \mathbb{R}$, and $\lambda_s$ is a factor used to avoid reallocating the buffer at every iteration.

\subsection{Tile-Primitive Mapping}
The goal of our mapping is to support two different access patterns with a
single intermediate buffer. During the forward pass, ray-primitive
intersections are discovered in a pixel-centric order. During the backward
pass, however, gradients are accumulated in a primitive-centric order: each
thread processes one primitive within a tile and gathers the contributions of
the pixels it covers. To make this access pattern efficient, we store intermediate values in a
primitive-major layout. For each primitive, entries are grouped by tile and then
by local pixel. This layout makes the backward reads contiguous for the thread
responsible for a given primitive-tile pair, while still allowing collision-free
writes during the forward pass.

\paragraph{Primitive offset}
During preprocessing, we compute a prefix sum over the projected footprints to construct a global offset table. For each primitive $i$, this table provides a base address in the contiguous buffer $\mathcal{LT}$.

\paragraph{$\mathcal{LT}$ Buffer access}
As illustrated in Figure~\ref{fig:forward_layout}, each ray-primitive intersection stores the accumulated
color $L$ and transmittance $T$ at a unique location in the $\mathcal{LT}$ buffer. The final address is obtained by decomposing the primitive footprint into a primitive base
offset, a tile-local offset, and a pixel-local offset:

\begin{figure}[h]
  \centering
  \includegraphics[width=\linewidth]{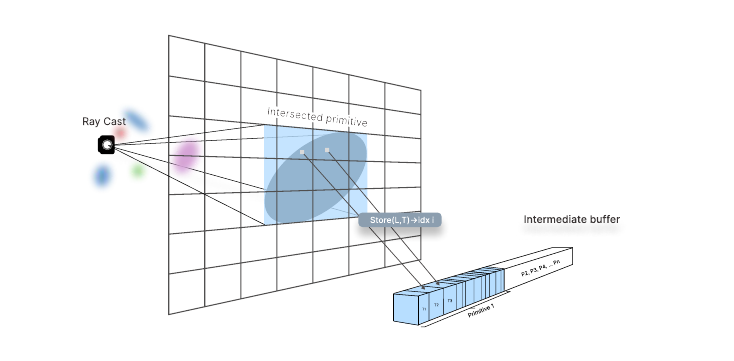}
  \caption{Forward pass. Each ray--primitive intersection stores the accumulated colour and transmittance tuple $L, T)$ in primitive-major tile-local order within our intermediate buffer.}
  \label{fig:forward_layout}
\end{figure}

\begin{equation}
idx_{global}(i, t, p) = \mathcal{O}_{primitive}(i) + \mathcal{O}_{tile}(i, t) + \mathcal{O}_{pixel}(i, t, p)
\end{equation}
where:
\begin{itemize}
    \item $\mathcal{O}_{primitive}$ is the primitive base offset in the $\mathcal{LT}$ Buffer;
    \item $\mathcal{O}_{tile} $ represents the cumulative number of pixels preceding the current tile $t$ within the $i$ primitive's AABB;
    \item $\mathcal{O}_{pixel}$ is the local index $p$ within the current tile of the primitive’s AABB.
\end{itemize}

\paragraph{Reverse Tile Mapping}

The forward indexing scheme provides collision-free writes, but the backward pass requires the inverse association: for each tile, we must know which
primitives overlap it. We construct this mapping during preprocessing.

\begin{itemize}
    \item Tile primitive: Each primitive emits one pair (tileID, primitiveID), for each tile overlapped by its AABB footprint. Then, these pairs are radix sorted by tileID, grouping all contributing primitives for each tile.
    \item Tile offset: Then, we build a tile-offset table by computing a prefix sum over the number of primitives per tile.
\end{itemize}

The backward pass can therefore retrieve the primitive list of any tile and its corresponding length.

\subsection{Primitive-centric backward propagation}

\begin{figure}[h]
  \centering \includegraphics[width=\linewidth]{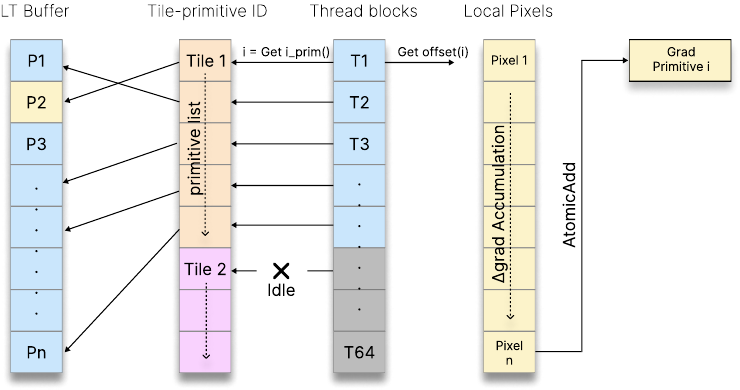}
  \caption{Backward pass: tile-particle mapping. Each thread iterates over a strided subset of primitives from the tile primitive list. For each primitive, it retrieves the corresponding $\mathcal{LT}$ buffer offset, accumulates gradients over the covered pixels, and writes the result to global memory using atomicAdd.}
  \label{fig:backward}
\end{figure}

Given the tile-primitive mapping, the backward pass launches one block per screen-space tile. 
As illustrated in Figure \ref{fig:backward}, each thread is assigned to one primitive overlapping that tile. \textit{Although assigning a full warp is possible, we found it less efficient in practice.} The thread then iterates over the pixels covered by the primitive within the tile, retrieves the corresponding $(L, T)$ values from $\mathcal{LT}$, and accumulates the local gradient contributions in registers.

After all covered pixels have been processed, the thread does an atomic update of the global primitive gradients. This changes the ownership of gradient accumulation:
Instead of letting pixel threads immediately scatter updates to shared primitive parameters, 3DGART accumulate contributions at the primitive tile
level before performing global atomic operations. This substantially reduces atomic contention and thread serialisation.

We use $8 \times 8$ tiles as a compromise between load balancing, aggregation overhead, and shared-memory usage. Smaller tiles reduce per-primitive pixel
variance, but increase the number of primitive-tile pairs and global accumulations. Larger tiles reduce aggregation overhead, but increase thread imbalance and require caching more per-pixel information, such as spherical-harmonic bases, which can exceed shared-memory limits and force additional per-pixel re-computation. 

\paragraph{Hybrid approach}Intermediate values $(L, T)$ are currently stored in \texttt{float}, resulting in a cost of 16 bytes per entry. In practice, most atomic contention arises from Spherical Harmonics (SH), accounting for 48 out of 59 atomic operations at order 3. This observation suggests a hybrid backward scheme, where colour gradients are accumulated using the particle-centric formulation, while the remaining terms are handled with a pixel-centric approach.

%% file: Chapters/5_experimentation.tex
\section{EXPERIMENTS AND ABLATION}

\subsection{Experimental setup}

\paragraph{Datasets.}
We evaluate 3DGART on standard novel-view synthesis benchmarks used by prior Gaussian rendering methods: Mip-NeRF 360 \cite{barron_mip-nerf_2021}, Tanks\&Temples \cite{Knapitsch2017}, and Deep Blending \cite{hedman_deep_2018}. 
Mip-NeRF 360 contains five outdoor scenes \textit{(Bicycle, Flowers, Garden, Stump, and Treehill)} and four indoor scenes \textit{(Bonsai, Counter, Kitchen, and Room)}, evaluated with downsampling factors of 4 and 2, respectively. For the global benchmark, we also evaluate on the \textit{Truck} and \textit{Train} scenes from Tanks\&Temples, as well as the \textit{DrJohnson} and \textit{Playroom} scenes from Deep Blending.

\paragraph{Approach.}
To evaluate efficiency, we implement three backward strategies: \textbf{Pixel}, the conventional per-pixel baseline; \textbf{Primitive}, our proposed tile-primitive approach accumulating gradients from a primitive-centric perspective; and \textbf{Hybrid}, which uses our primitive formulation for color gradients while retaining the pixel-centric scheme for remaining terms.

\paragraph{Hardware and metrics.}
All experiments are conducted on the same NVIDIA RTX 4090 GPU.
Training times are measured as wall-clock time, excluding image-loading overhead. We report PSNR, SSIM, and LPIPS using a standardised evaluation pipeline across all methods. For LPIPS, we follow the evaluation convention used in 3DGS. 

\subsection{Raw Training Speed Benchmark}
To evaluate 3DGART's backward efficiency, we perform a raw head-to-head benchmark between a common per-pixel and our tile-primitive approach. We do not use a maximum ray intersection cap, we set the Gaussian kernel to $k=2$ (as 3DGS), and $\alpha_{\min}=1/255$. We evaluate both methods increasing primitive budgets to measure scalability.

\begin{table}[h]
    \centering
    \small
    \caption{
    Raw head-to-head training-speed benchmark on Mip-NeRF 360.
    Gain is measured against the corresponding pixel-based representation.
    }
    \label{tab:raw_training_speed}
    \renewcommand{\arraystretch}{1.45}
    \setlength{\tabcolsep}{12pt}

    \begin{tabular}{
        c
        !{\vrule width 0.6pt}
        ccc
        c
    }
        \Xcline{2-5}{0.45pt}
        &
        \multicolumn{4}{c}{\textbf{Mip-NeRF 360}} \\
        \Xcline{1-5}{0.45pt}

        \textbf{\#G}
        & \textbf{Pixel}
        & \textbf{Hybrid}
        & \textbf{Primitive}
        & \textbf{Gain}$\uparrow$ \\
        \Xhline{0.5pt}

        1.00M
        & 44m08s
        & 19m55s
        & \cellcolor{highlightgreen!35}12m35s
        & \cellcolor{highlightgreen!35}3.51$\times$ \\

        2.00M
        & 51m13s
        & 24m56s
        & \cellcolor{highlightgreen!15}16m32s
        & \cellcolor{highlightgreen!15}3.10$\times$ \\

        \Xhline{1.0pt}
    \end{tabular}
\end{table}

\begin{table*}[t]
    \centering
    \small
    \caption{
    Global benchmark across Mip-NeRF 360, Tanks\&Temples, and Deep Blending. Our approach significantly reduces training time over the current state-of-the-art fully ray tracing approach (3DGRT), while matching rasterisation-based methods.
    }
    \label{tab:global_benchmark}
    \renewcommand{\arraystretch}{1.65}
    \newcommand{\datasetvrule}{!{\color{black!45}\vrule width 0.45pt}}
    \newcommand{\groupsep}{\Xhline{0.45pt}}
    \newcommand{\thinsep}{\Xhline{0.25pt}}
    \setlength{\tabcolsep}{3pt}

    \begin{tabular}{
        llc
        !{\vrule width 0.6pt}
        ccccc
        !{\vrule width 0.6pt}
        ccccc
        !{\vrule width 0.6pt}
        ccccc
    }
        \Xcline{4-18}{0.45pt}
        & & &
        \multicolumn{5}{c}{\textbf{Mip-NeRF 360}} &
        \multicolumn{5}{c}{\textbf{Tanks\&Temples}} &
        \multicolumn{5}{c}{\textbf{Deep Blending}} \\
        \Xcline{1-18}{0.45pt}

        \textbf{Model} & \textbf{Type} & \textbf{Iter}
        & \textbf{PSNR}$\uparrow$
        & \textbf{SSIM}$\uparrow$
        & \textbf{LPIPS}$\downarrow$
        & \textbf{Train}
        & \textbf{\#G}
        & \textbf{PSNR}$\uparrow$
        & \textbf{SSIM}$\uparrow$
        & \textbf{LPIPS}$\downarrow$
        & \textbf{Train}
        & \textbf{\#G}
        & \textbf{PSNR}$\uparrow$
        & \textbf{SSIM}$\uparrow$
        & \textbf{LPIPS}$\downarrow$
        & \textbf{Train}
        & \textbf{\#G} \\
        \Xhline{0.5pt}

        ZipNeRF & -- & --
        & 28.54 & 0.828 & 0.219 & 5h & --
        & -- & -- & -- & -- & --
        & -- & -- & -- & -- & -- \\

        \Xhline{0.45pt}

        3DGS & Rast. & 30K
        & 27.35 & 0.814 & 0.216 & 20m19s & 3.31M
        & \cellcolor{highlightgreen!15}23.52 & 0.842 & 0.183 & 11m53s & 1.84M
        & 29.77 & 0.906 & 0.263 & 20m55s & 2.81M \\

        3DGUT & Rast. & 30K
        & \cellcolor{highlightgreen!15}27.47 & 0.812 & 0.218 & 16m09s & 3.23M
        & \cellcolor{highlightgreen!35}23.52 & \cellcolor{highlightgreen!15}0.848 & 0.173 & 11m24s & 2.15M
        & 29.60 & 0.902 & 0.256 & 8m35s & 1.37M \\

        HTGS & Rast. & 30K
        & 27.04 & \cellcolor{highlightgreen!15}0.818 & \cellcolor{highlightgreen!35}0.196 & 10m09s & 2.20M
        & 23.09 & 0.846 & 0.172 & 6m32s & 0.77M
        & 29.93 & 0.907 & \cellcolor{highlightgreen!35}0.234 & 8m25s & 1.01M \\

        \Xhline{0.45pt}

        3DGRT & RT & 30K
        & 27.11
        & 0.809
        & 0.214
        & 62m21s
        & 4.19M
        & 22.81 & 0.844 & \cellcolor{highlightgreen!35}0.165 & 36m47s & 4.05M
        & 29.62 & 0.904 & 0.247 & 50m51s & 1.74M \\

        GRay & RT & 15K
        & 26.79
        & 0.799
        & \cellcolor{highlightgreen!15}0.198
        & 7m21s
        & 1.68M
        & 22.38 & 0.821 & \cellcolor{highlightgreen!15}0.170 & 3m58s & 1.04M
        & 29.18 & 0.891 & \cellcolor{highlightgreen!15}0.238 & 5m28s & 1.21M \\

        \Xhline{0.45pt}

        \textbf{Ours} & RT & 30K
        & 27.31
        & 0.805
        & 0.239
        & 12m35s
        & 1.00M
        & 23.20
        & \cellcolor{highlightgreen!35}0.852
        & 0.175
        & 7m54s
        & 1.00M
        & \cellcolor{highlightgreen!35}30.20
        & \cellcolor{highlightgreen!35}0.913
        & 0.242
        & 13m06s
        & 1.00M \\

        \textbf{Ours} & RT & 30K
        & \cellcolor{highlightgreen!35}27.51
        & \cellcolor{highlightgreen!35}0.820
        & 0.211
        & 16m32s
        & 2.00M
        & -- & -- & -- & -- & --
        & -- & -- & -- & -- & -- \\


        \Xhline{1.0pt}
    \end{tabular}
\end{table*}

Table \ref{tab:raw_training_speed}, 3DGART achieves a mean gain factor of $3.51\times$ and $3.1\times$ with $1.0  \times  10^6$ and $2.0  \times  10^6$ primitives, respectively, over the baseline per-pixel implementation. We observe that the performance gain tends to decrease as the number of primitives grows. As primitives shrink, the number of intersections per primitive decreases, which naturally reduces the atomic contention.

\subsection{Micro Benchmarks}

\begin{figure}[h]
    \centering
    \includegraphics[width=\linewidth]{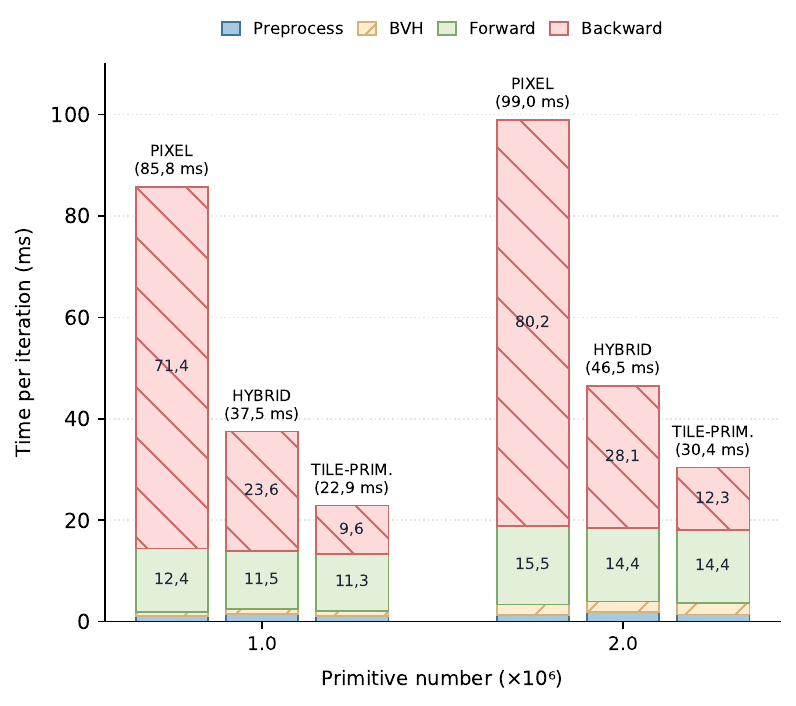}
    \caption{\textbf{Per-component iteration time (ms).} Runtime breakdown for the per-pixel baseline, our hybrid approach, and our tile-primitive method at $1.0\times10^6$ and $2.0\times10^6$ primitives, highlighting the reduction in backward pass overhead.}
    \label{fig:micro_benchmark}
\end{figure}

To better understand where this end-to-end acceleration comes from, we analyse each pipeline component. Specifically, the pre-processing includes per-primitive instance transform computation, our tile-primitive and additional forward pre-computations.
(e.g., $w2g=S^{-1}R^\top$ and $o_g$), followed by BVH build/update, forward rendering, and backward propagation.

As illustrated in Figure \ref{fig:micro_benchmark}, the backward pass remains the major bottleneck in conventional per-pixel approach, representing 81-86\% of the iteration time. In contrast, our architecture makes the backward pass comparable to, or even faster than, the forward pass. Compared to per-pixel backpropagation, our tile-primitive method is 7.3$\times$ and 6.5$\times$ faster at 1.0 and 2.0 $\times 10^6$ primitive, respectively.

Since forward rendering remains dominated by per-pixel ray tracing and scales primarily with image resolution, matching its latency in the backward pass indicates that the atomic contention typically found in ray-traced Gaussian training has been largely mitigated by our primitive-based approach.

Our hybrid approach reduces VRAM consumption by a factor of $4$ but partially reintroduces serialized atomic contention. As a result, training is approximately $1.5\times$ slower than with our tile-primitive approach.

\subsection{Memory Analysis}

\paragraph{Resolution and primitive dependent footprint} 
To assess scalability, we analyse the VRAM consumption of the $LT$ buffer on Mip-NeRF 360 as the number of primitives increases. We progressively increase the number of Gaussians by $1.0 \times 10^6$ and report two resolution settings (Low, High). Outdoor scenes are downsampled by factors of 8 and 4, while indoor scenes are downsampled by factors of 4 and 2, respectively.

\begin{figure}[h]
    \centering
    \includegraphics[width=\linewidth]{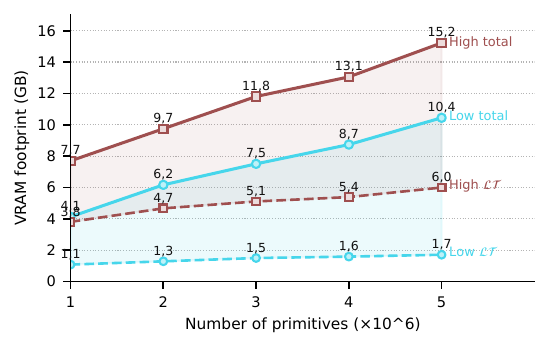}
    \caption{\textbf{Peak memory analysis.} Comparison of total VRAM usage and $\mathcal{LT}$ buffer consumption across primitive scales ($1\text{M}$ to $5\text{M}$) for our tile--primitive approach at low and high resolutions (up to $1.5\text{ MP}$). For the hybrid approach, the low and high settings correspond to resolutions up to $1.5\text{ MP}$ and $6\text{ MP}$, respectively.}
    \label{fig:vram}
\end{figure}

Since our approach relies on perspective-correct AABB footprints, the size of the $\mathcal{LT}$ buffer scales linearly with the total number of pixels. As shown in Figure~\ref{fig:vram}, the $\mathcal{LT}$ buffer memory usage also increases approximately linearly with the number of primitives. Nevertheless, for the resolutions considered in our benchmarks (up to approximately 1.5 megapixels), the additional memory remains well within the capacity of modern GPUs, making the tile-primitive approach practical for standard training workloads.

For higher-resolution settings, where the $\mathcal{LT}$ buffer may become the limiting factor, our hybrid backward strategy provides an alternative by reducing $\mathcal{LT}$ VRAM consumption by a factor of 4. We therefore view the hybrid formulation as a practical trade-off for memory-constrained or high-resolution scenarios.

Importantly, this overhead is strictly confined during training; at inference time, our method preserves the original memory footprint of standard Gaussian ray tracing models.

\subsection{Global benchmark}

We further evaluate 3DGART against rasterisation-based (RAST) and ray-tracing-based (RT) methods in Table~\ref{tab:global_benchmark}. In this comparison, 3DGS denotes its modern pre-trained model, 3DGUT denotes its unsorted variant, and HTGS denotes the state-of-the-art efficient ray-based rasterizer, evaluated with anti-aliasing enabled. 

3DGRT enabled fully ray-traced pipeline, as well as the uses of secondary rays, but introduced several implementation compromises to both accelerate ray tracing and reduce atomic contention during the backward pass, including an $\alpha_{\min}$ threshold, a capped number of ray--primitive intersections per pixel, and narrow Gaussian kernels (Increasing the number of primitives to represent a scene). GRay introduces additional compromises through its dense initialization strategy, which can exacerbate aliasing artifacts. Consequently, it achieves an impressive speed while trading-off PSNR/SSIM Scores. 

In contrast, 3DGART directly addresses the source of the contention bottleneck through a primitive-centric backward formulation. Instead of relying on heuristic approximations, our method preserves the original alpha compositing formulation while achieving practical training efficiency and high reconstruction fidelity, as demonstrated in Figure~\ref{fig:teaser}.

%% file: Chapters/6_discussion.tex
\section{Limitations and future work}

\paragraph{Complex cameras} As mentioned, we utilise perspective-correct projection; however, this approach is inherently limited to perspective camera models and does not support extreme wide-angle or non-linear projections such as fisheye lenses. To overcome this, future work could replace this projection with an exact projective geometry formulation \cite{huang_3dgeer_2026}.

\paragraph{VRAM Consumption.}
Our approach trades VRAM consumption for significantly faster gradient accumulation. Although our hybrid backward formulation reduces memory usage by a factor of 4, making high-resolution training (e.g., 4K) practical, it partially sacrifices the performance benefits of our tile-primitive approach. While storing the $\mathcal{LT}$ buffer in \texttt{float16} provides a straightforward way to further reduce memory consumption, we believe more principled solutions deserve investigation to preserve the fully primitive-centric backward formulation.

%% file: Chapters/7_conclusion.tex
\section{Conclusion}

We presented 3D Gaussian Accelerated Ray Tracing (3DGART), a framework that addresses the primary computational bottleneck of differentiable Gaussian ray tracing: atomic contention in the backward pass. By shifting from a pixel-centric to a particle-centric formulation, our tile-based architecture assigns each CUDA thread exclusive ownership over gradient accumulation within a tile, transforming a contention-heavy scatter operation into a structured gather-like computation. While trading off VRAM for training speed, our method achieves an $\approx 4\times$ speedup over 3DGRT on the Mip-NeRF 360 dataset while simultaneously improving reconstruction quality. To maintain efficiency at higher resolutions, we further design a hybrid approach. Overall, 3DGART demonstrates that fully ray-traced Gaussian training becomes practical when gradient accumulation is structured around primitives rather than pixels.

%% file: Chapters/acknowledgement.tex
\begin{acks}

I gratefully acknowledge the financial support provided by the New Zealand Government.  I am especially grateful to Oliver Batchelor for his guidance and for helping shape and refine the ideas presented in this paper. I would also like to thank Florian Hahlbohm and Nicolas Moenne-Loccoz for their valuable assistance with the mathematical foundations underlying this work. Finally, I thank Richard Green for his supervision, support, and continued encouragement.

\end{acks}